\documentclass[epj,referee]{svjour}
\usepackage{graphics}
\begin{document}
\title{Microwave conductivity of heavy fermions in UPd$_{2}$Al$_{3}$}
\author{Marc Scheffler\inst{1} \and Martin Dressel\inst{1} \and Martin Jourdan\inst{2}% etc
}                     % Do not remove
\institute{1.~Physikalisches Institut, Universit\"at Stuttgart, D-70550 Stuttgart, Germany \and Institut f\"ur Physik, Johannes Gutenberg Universit\"at, D-55099 Mainz, Germany}
\date{Received: date / Revised version: date}
\abstract{
Heavy-fermion compounds are characterized by electronic correlation effects at low energies which can directly be accessed with optical spectroscopy. Here we present detailed measurements of the frequency- and temperature-dependent conductivity of the heavy-fermion compound UPd$_{2}$Al$_{3}$ using broadband microwave spectroscopy in the frequency range 45~MHz to 40~GHz at temperatures down to 1.7~K. We observe the full Drude response with a relaxation time up to 50~ps, proving that the mass enhancement of the heavy charge carriers goes hand in hand with an enhancement of the relaxation time. We show that the relaxation rate as a function of temperature scales with the dc resistivity. We do not find any signs of a frequency-dependent relaxation rate within the addressed frequency range.
 % end of PACS codes
} %end of abstract
\maketitle

\section{Introduction}
\label{intro}
Heavy-fermion materials are intermetallic compounds that exhibit peculiar electronic properties at low temperatures, such as high values of the specific heat and the magnetic susceptibility.\cite{Grewe1991} These effects are explained with an effective mass of the mobile charge carriers that is enhanced up to 1000 times with respect to the free electron mass. In the Kondo picture, the origin of this high effective mass is the hybridization of different sets of electrons, namely of conventional metallic band electrons and of barely localized f-electrons introduced by elements like Ce, Yb, or U that are part of these compounds. The hybridization leads to \lq barely mobile\rq{} electrons that cause the characteristic heavy-fermion properties. All these effects are restricted to low energies; i.e.\ they only occur at low temperatures (a typical temperature range is below 10~K), and they can only be studied with low-energy probes.
Here optical spectroscopy \cite{Dressel2002a,Degiorgi1999} is particularly suited: firstly, the electromagnetic radiation directly couples to the electric charges, i.e.\ the heavy fermions themselves. Secondly, the energy of the probe, i.e.\ the frequency of the employed radiation, can be tuned over many orders of magnitude to match the processes of interest.
This need of optical spectroscopy at extremely low frequencies (GHz and THz) compared to conventional optics has lead to the unsatisfying situation that of the two fundamental questions in heavy-fermion physics that have been the focus of the respective optical experiments, namely the possible presence of a hybridization gap and the dynamics of the heavy charge carriers, only the first one could be fully addressed whereas the second remained unsolved due to the experimental difficulties. 
Only recently we were able to combine high-quality thin film samples \cite{Huth1993,Huth1994} and a new spectrometer \cite{Scheffler2005a} to obtain broadband microwave conductivity spectra that revealed the full charge dynamics of a heavy-fermion compound, UPd$_2$Al$_3$.\cite{Scheffler2005b,Scheffler2005c} In this article we describe these experiments in detail and discuss their implications for our understanding of heavy fermions.

\subsection{Drude response of metals}

Heavy-fermion materials are metals, and their optical response is discussed within the same framework as normal metals, that is the Drude response.
Following the original model of Paul Drude,\cite{Drude1900} the frequency-dependent conductivity $\sigma(\omega)=\sigma_1(\omega) + i \sigma_2(\omega)$ of a metal is described within a relaxation approach:\cite{Dressel2002a}
\begin{equation}\label{EqFreqDepCond}
  \sigma(\omega) = \sigma_{0} \frac{1}{1-i\omega\tau} 
  	= \sigma_{0} \left( \frac{1}{1+\omega^2 \tau^2} + i \frac{\omega \tau}{1+\omega^2 \tau^2} \right) \, ,
\end{equation}
where $\sigma_{0} = \frac{ne^{2}\tau}{m}$ is the dc conductivity, $\tau$ the relaxation time, $\omega = 2 \pi f$ the angular frequency, $e$ the elementary charge, $n$ the density of mobile electrons, and $m$ their effective mass. This formula already highlights the importance of the relaxation time and the effective mass. The mentioned expression for $\sigma_0$ originates from the \textit{classical} Drude model and cannot be applied to actual metallic electrons that have to be described using quantum mechanics. But the frequency dependence of Eq.\ (\ref{EqFreqDepCond}) only relies on the assumption of an exponential relaxation of the mobile charges and therefore is fully consistent with quantum mechanics.\cite{Dressel2002a,Dressel2006}

\subsection{Optical properties of heavy fermions}

A hypothetical direct transition from a metallic, non-inter\-acting electron system to the interacting, heavy-fermion state (a transition tuned purely by increasing the interaction) is characterized by strong modifications of effective mass $m^*$ and relaxation time $\tau ^*$; their renormalization is expected to scale:\cite{Varma1985,Millis1987}
\begin{equation}\label{EqMTauScaling}
  m^*/m = \tau^*/\tau \, ,
\end{equation}
i.e. the strongly enhanced mass goes hand in hand with an increase of the relaxation time. In fact the two are just different manifestations of the same effect: a large effective mass is equivalent to a high density of states at the Fermi level, which in turn is equivalent to a low Fermi velocity which means that the scattering rate is low (if the mean free path is unaffected, e.g.\ in the case of impurity scattering) and the relaxation time long.

The relaxation rate $\Gamma=1/\tau$, for usual metals in the infrared frequency range, here is expected to shift to the THz and microwave frequency ranges. This shift of the Drude response to very low frequencies is the signature of heavy-fermion behavior in the electrical conductivity whereas the absolute value of the dc conductivity is similar to that of normal metals and does not indicate heavy-fermion behavior.
To avoid confusion later on, we will always employ starred quantities to describe the actual, renormalized behavior of the heavy-fermion state; i.e.\ Eq.\ (\ref{EqFreqDepCond}) then reads
\begin{equation}\label{EqFreqDepCondRenorm}
  \sigma(\omega) = \sigma_0^* \; \frac{1}{1-i\omega\tau^*} \, .
\end{equation}

The extremely low frequency of the Drude roll-off is the first fundamental expectation for the optical conductivity of heavy fermions and has been addressed experimentally using microwave cavity resonators, but these studies could not reveal the full frequency dependence.\cite{Sulewski1988,Beyermann1988a,Awasthi1993,Degiorgi1997,Tran2002,Scheffler2009}
A second feature in the optical response of heavy fermions is usually termed \lq hybridization gap\rq{} and has been studied in numerous heavy-fermion materials, but occurs at higher frequencies than relevant for the present study.\cite{Dordevic2001,Singley2002,Mena2005,Okamura2007}

\subsection{Frequency-dependent relaxation rate}\label{FreqDepRelRate}

Eq.\ (\ref{EqFreqDepCondRenorm}) assumes a relaxation rate (and an effective mass) that is independent of frequency. However, for heavy fer\-mi\-ons a frequency-dependent relaxation rate can occur.
This frequency dependence is obtained from experimental data via the extended Drude formalism.\cite{Dressel2002a} Here, Eq.\ (\ref{EqFreqDepCondRenorm}) is modified by employing explicitly frequency-dependent effective mass $m^*(\omega)$ and relaxation time $\tau^*(\omega)$. Furthermore, the definition of the unrenormalized plasma frequency $\omega_p = \sqrt{4 \pi n e^2 / m_0}$ with $m_0$ the bare electron mass is used to obtain the frequency dependences of the mass enhancement $m^*/m_0$ and the unrenormalized relaxation rate $\Gamma = 1/ \tau$:
\begin{equation}\label{EqFreqDepGamma}
  \Gamma(\omega) = \frac{1}{\tau(\omega)} = \frac{\omega_p^2}{4 \pi} \rm{Re}\left\{\frac{1}{\sigma(\omega)}\right\}
                 = \frac{\omega_p^2}{4 \pi} \frac{\sigma_1(\omega)}{|\sigma(\omega)|^2}
                 = \frac{\omega_p^2}{4 \pi} \rho_1(\omega) \, ,
\end{equation}
\begin{equation}\label{EqFreqDepMstar}
  \frac{m^*}{m_0} =  -\frac{\omega_p^2}{4 \pi \omega} \rm{Im}\left\{\frac{1}{\sigma(\omega)}\right\}
                 = \frac{\omega_p^2}{4 \pi \omega} \frac{\sigma_2(\omega)}{|\sigma(\omega)|^2}
                 = -\frac{\omega_p^2}{4 \pi \omega} \rho_2(\omega) \, ,
\end{equation}
where $\rho(\omega) = \rho_1 + i \rho_2 = 1/\sigma(\omega)$ was used. 
Applying Eqs.\ (\ref{EqFreqDepGamma}) and (\ref{EqFreqDepMstar}) to experimentally obtained conductivity data $\sigma(\omega)$ requires knowledge of the unrenormalized plasma frequency $\omega_p$. For simple metals, this quantity can be calculated if the charge carrier density $n$ is known, or it can be obtained from optical measurements at higher frequencies (and higher temperatures). However, the plasma frequency of heavy fermions cannot be determined easily, and in particular one cannot assume \textit{a priori} that it is independent of temperature.

\subsection{Heavy-fermion material UPd$_2$Al$_3$}

The material we have studied, UPd$_2$Al$_3$, is a prominent heavy-fermion compound:\cite{Geibel1991,Jourdan1999,Sato2001} its superconducting critical temperature $T_c$ = 2.0~K is rather high for heavy fermions, and the superconductivity develops within an antiferromagnetic state ($T_N \approx$ 14~K). For the current study, we focus on the heavy-fermion state above $T_c$, and the superconducting state at lower temperatures is only important for the calibration procedure.

%\begin{figure}
%\includegraphics[width=8cm]{Fig1_SamplesRdc.eps}
%\caption{\label{FigSamplesRdc}(Color online) Main plot: temperature dependence of dc resistivity for samples 1, 2, and 3. Data between 1.7~K and 300~K was obtained in 2-point geometry with the microwave Corbino measurements, during cooldown before the microwave experiments were performed. Features around 5~K and 130~K are artifacts of these 2-point measurements. The additional data down to 0.3~K for sample 2 was obtained with an independent 4-point measurement in a $^3$He cryostat. Inset: low-temperature resistivity of sample 1 obtained in 4-point geometry, indicating $T^2$-behavior.}
%\end{figure}

\begin{figure}
\resizebox{1.00\columnwidth}{!}{%
  \includegraphics{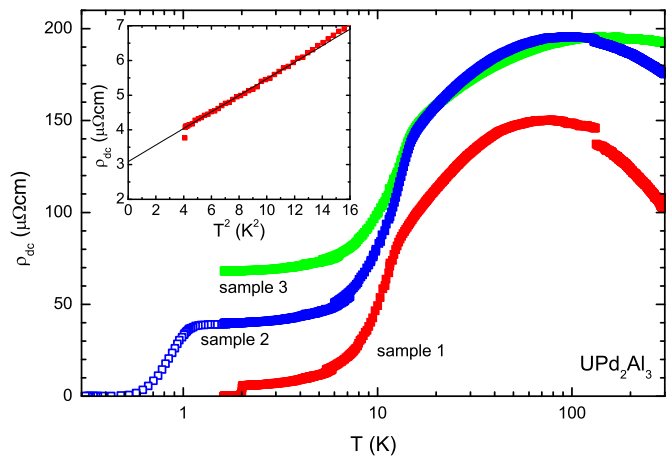}
}
%\vspace{5cm}       % Give the correct figure height in cm
\caption{(Color online) Main plot: temperature dependence of dc resistivity for UPd$_2$Al$_3$ samples 1, 2, and 3. Data between 1.7~K and 300~K was obtained in 2-point geometry with the Corbino setup, during cooldown before the microwave experiments were performed. Features around 5~K and 130~K are artifacts of these 2-point measurements. The additional data down to 0.3~K for sample 2 was obtained with an independent 4-point measurement in a $^3$He cryostat. Inset: low-temperature resistivity of sample 1 obtained in 4-point geometry, indicating $T^2$-behavior.}
\label{FigSamplesRdc}       % Give a unique label
\end{figure}

UPd$_2$Al$_3$ crystallizes in a hexagonal structure and exhibits typical heavy-fermion behavior with an effective mass of $m^*=66 \, m_0$ as determined from thermodynamic measurements.\cite{Geibel1991} The temperature dependence of the dc conductivity is shown in Fig.\ \ref{FigSamplesRdc} for different samples. At high temperatures the resistivity does not depend strongly on temperature: cooling from 300~K the conductivity first increases slightly; a common feature for heavy fermions. Below approximately 100~K, the resistivity decreases slowly, followed by a much steeper decrease below the N\'eel temperature. Toward even lower temperatures, the temperature dependence of the resistivity flattens due to residual scattering, until superconductivity sets in.

The optical properties of UPd$_2$Al$_3$ at frequencies above 1~cm$^{-1}$ = 30~GHz have already been investigated in detail from the ultraviolet down to the THz range.\cite{Degiorgi1997,Dressel2002b,Dressel2002c} The previous studies on UPd$_2$Al$_3$ found typical heavy-fermion behavior with a hybridization-gap-like feature (a minimum in $\sigma_1 (f)$ around $f/c = $100~cm$^{-1}$). But even studies down to 1~cm$^{-1}$ could not observe the Drude roll-off; instead an additional maximum in $\sigma_1 (f)$ was found around 4~cm$^{-1}$ at temperatures below $T_N$ and was interpreted as a signature of a pseudogap induced by the antiferromagnetic state.\cite{Dressel2002b,Dressel2002c} Previous attempts to resolve the low-temperature conductivity of UPd$_2$Al$_3$ at frequencies below 1~cm$^{-1}$ used cavity resonators, but they remained inconclusive.\cite{Degiorgi1997,Dressel2002b,Dressel2002c}

\section{Experiment}

Our broadband microwave spectrometer \cite {Scheffler2005a} is most sensitive to heavy-fermion samples when they are thin films.\cite{RemarkSensitivityBulk} 
We have grown thin films of UPd$_2$Al$_3$ using MBE techniques: evaporation of the three constituents U, Pd, Al and deposition onto LaAlO$_3$(111) substrates.\cite{Huth1993,Huth1994,Jourdan1999} The excellent quality of these thin films is evident from x-ray analysis of the lattice as well as from the temperature dependence of the dc resistivity, which reproduces the features known from bulk single crystals.

The microwave experiments were performed with a broadband spectrometer that covers the frequency range 45~MHz\ --\ 40~GHz and temperatures from 300~K down to 1.1~K.\cite{Scheffler2005a,Steinberg2008} This spectrometer is based on a Corbino probe: the thin film sample is pressed flat against the open end of a coaxial cable and thus reflects the microwave signal that travels in the cable. The reflection coefficient (which directly reveals the sample impedance) is measured by a commercial vector network analyzer at room temperature, whereas the Corbino probe is located in a purpose-designed $^4$He cryostat. Crucial requirement for reproducible measurements is a reliable calibration which corrects for those contributions to the reflection coefficient that do not stem from the sample (e.g. damping and phase shift in the coaxial cable). Here we can employ two different procedures:\cite{Scheffler2005a} on the one hand a three-standard calibration with known calibration samples for open (teflon), short, and load (NiCr thin film) and on the other hand a short-only calibration with just one calibration measurement. We have shown \cite{Scheffler2005a} that for low-impedance samples (like the UPd$_2$Al$_3$ samples discussed here) these two procedures lead to equivalent results. Furthermore, we can use two different short standards: either a conventional bulk metal (aluminum) or the sample under study itself, if it becomes superconducting at temperatures slightly lower than those of interest. One sample of the present study was part of the experimental proof that also these two procedures give the same general result.\cite{Scheffler2005a} In addition to being more convenient, the calibration with the superconducting sample also leads to smaller errors in the finally obtained quantity, i.e.\ the impedance or conductivity. Unfortunately this latter procedure only works for rather low temperatures where the transmission properties of the coaxial cable do not change with the increasing temperature during a measurement. For the present study, this holds up to approximately 15 K. Using the superconducting calibration we obtain reliable data up to 40~GHz whereas with the non-superconducting calibration we are usually limited to frequencies up to 20~GHz.

The calibration is performed separately for each frequency and each temperature. From the complex reflection coefficient obtained after applying the calibration, we directly calculate the complex conductivity.\cite{Scheffler2005a} This is particularly simple due to the thin film samples: in the complete temperature and frequency range of the present study, the film thickness is much smaller than the skin depth, and therefore we can assume that the fields (and current density) are uniform within the film thickness.\cite{RemarkAnomalousSkinEffect}

A reliable calibration can only be achieved if the temperature distribution along the coaxial cable in the cryostat is exactly reproduced for sample and calibration measurements. We therefore employ a strict operating procedure for the spectrometer, and we use computer control for the actual measurements at numerous temperatures (in the present study: 111 temperature points) during the warming from 1.65~K to 300~K. Thus, for any low-temperature measurement we obtain the full temperature dependence up to room temperature.

In addition to the microwave experiments, we simultaneously measure the dc resistance. This characterizes the samples, proves their high quality, and is direct evidence for small contact resistance between sample and probe. Good contact is essential as our two-point microwave im\-ped\-ance measurements cannot distinguish between contributions of the sample and the contact resistance. The measured dc resistance also serves as in-situ sensor to determine the actual sample temperature.\cite{Scheffler2005a}

\section{Results}

\begin{figure}
\resizebox{1.00\columnwidth}{!}{%
  \includegraphics{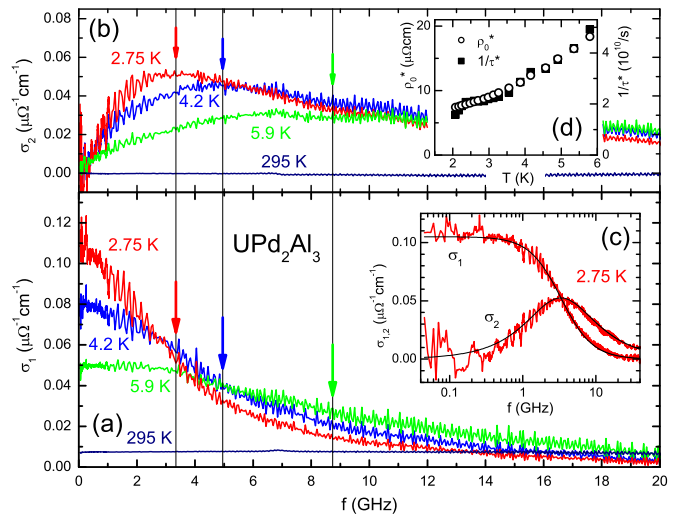}
}
%\vspace{5cm}       % Give the correct figure height in cm
\caption{(Color online) Microwave spectra of the real part $\sigma_1$ [panel (a)] and the imaginary part $\sigma_2$ [panel (b)] of UPd$_2$Al$_3$ for several temperatures and frequencies up to 20~GHz. At high temperatures, $\sigma_1$ is constant and $\sigma_2$ vanishes. For low temperatures, clear Drude features dominate the spectra: roll-off in $\sigma_1$ and maximum in $\sigma_2$. (The regular oscillations are traces of standing waves in our setup that are not completely taken account for by the calibration procedure.) Inset (c) shows the exemplary spectrum at 2.75~K on a logarithmic frequency scale (covering the complete range 45~MHz to 40~GHz) together with a combined Drude fit, Eq.\ (\ref{EqFreqDepCondRenorm}), of real and imaginary parts. Inset (d) displays the temperature dependence of $\rho_0^* = 1/\sigma_0^*$ and $1/\tau_0^*$, as deduced from the Drude fits to the spectra at different temperatures.}
\label{FigSpectraEtc}       % Give a unique label
\end{figure}

\subsection{Microwave conductivity and Drude response}

In panels (a) and (b) of Fig.\ \ref{FigSpectraEtc} we show the real part $\sigma_1$ and the imaginary part $\sigma_2$, respectively, of the microwave conductivity of UPd$_2$Al$_3$ for a set of different temperatures. For all temperatures and for frequencies below 500~MHz, $\sigma_1$ is almost constant (more obvious from the logarithmic plot in inset (c) of Fig.\ \ref{FigSpectraEtc}) and matches the dc conductivity that was determined independently (compare Fig.\ \ref{FigSamplesRdc}). This coincidence of dc and low-frequency microwave conductivity indicates that the conductivity is frequency-independent at all frequencies below our range; i.e. all relevant dynamical properties of the electrons in UPd$_2$Al$_3$ occur at frequencies above 500~MHz.

At temperatures above 10~K, $\sigma_1$ as a function of frequency is basically flat and $\sigma_2$ is zero in the whole observed frequency range (as an example, data for $T$~= 295~K is shown in Fig.\ \ref{FigSpectraEtc}), indicating that the relaxation rate at these temperatures is much higher than the frequencies addressed here. This is consistent with general expectations for the relaxation rate of a metal and with the previous optical measurements.\cite{Dressel2002a,Degiorgi1997} At lower temperatures, a roll-off in $\sigma_1$ clearly develops together with a concomitant maximum in $\sigma_2$; both move toward lower frequencies with decreasing temperature. These features are the Drude response of the heavy fermions that previously eluded direct observation and is the focus of the present study. 
In this experiment, we observe the transition from diffusive ($\omega < 1/\tau^*$) to ballistic ($\omega > 1/\tau^*$) transport as a function of frequency within the low-temperature spectra and as a function of temperature (for fixed frequency); the latter is evident from the reversed order of the $\sigma_1$-spectra at low frequencies compared to high frequencies: the diffusive-to-ballistic transition is characterized by a maximum in $\sigma_1$ versus temperature.\cite{Scheffler2005c}

In Inset (c) of Fig.\ \ref{FigSpectraEtc} we show the real and imaginary parts of the microwave conductivity of UPd$_2$Al$_3$ at a representative low temperature of 2.75~K, together with a combined fit following the Drude formula Eq.\ (\ref{EqFreqDepCondRenorm}). This fit with only two free parameters, $\sigma_0^*$ and $\tau^*$, gives an excellent description of the data in real \textit{and} imaginary parts, thus establishing these experiments as the prime examples of simple Drude behavior. Since $\sigma_0^*$ has to coincide with the dc conductivity, the fit in fact only has one previously unknown fit parameter, namely the relaxation time.

The Drude fit applies to the full frequency range, and in particular there is no increase observed in $\sigma_1(f)$ even for frequencies approaching 40~GHz. This observation is noteworthy because for correlated electron systems one might expect charge carriers with different effective masses and relaxation rates within the same material. If there were charge carriers present with a considerably higher relaxation rate, they should contribute to a finite value of $\sigma_1(f)$, but we do not see any such signs. Futhermore, an increase in $\sigma_1(f)$ for somewhat higher frequencies is expected from previous THz studies on UPd$_2$Al$_3$.\cite{Dressel2002b,Dressel2002c}

The value for the relaxation time, here $\tau^* = 4.8 \times 10^{-11}$~s for $T = 2.75$~K, can be compared with results from de Haas-van Alphen studies,\cite{Shoenberg1984} where Dingle temperatures between 0.10~K and 0.28~K were found for different orbits in UPd$_2$Al$_3$ at even lower measurement temperatures,\cite{Inada1999} corresponding to relaxation times from $4.3\times 10^{-12}$~s to $1.2\times 10^{-11}$~s. Such a discrepancy, with the relaxation time obtained from transport being a factor 10--100 longer than those from de Haas-van Alphen studies, is commonly observed.\cite{Shoenberg1984}
The low-temperature relaxation rate that we find for UPd$_2$Al$_3$ is smaller than those of the heavy-fermion compounds CePd$_3$ and CeAl$_3$ as determined previously from microwave cavity experiments.\cite{Beyermann1988a,Awasthi1993} This can be explained by the high quality of our sample (evident from the small residual resistivity) and the corresponding weak impurity scattering, which is the dominant scattering mechanism at low temperatures.

The conductivity spectra in panels (a) and (b) of Fig.\ \ref{FigSpectraEtc} show how the characteristic roll-off in $\sigma_1(\omega)$ and the coincident maximum in $\sigma_2(\omega)$, which occur at the relaxation rate, shift to higher frequencies when the temperature is raised. This increase in relaxation rate is expected and is due to the more frequent scattering events at higher temperature.
The temperature dependence of the relaxation rate (obtained from Drude fits to the conductivity spectra) is shown in inset (d) of Fig.\ \ref{FigSpectraEtc}, together with the dc resistivity $\rho_0^* = 1/\sigma_0^*$ (also obtained from the Drude fits). As can clearly be seen, the dc resistivity and the relaxation rate have the same temperature dependence in the temperature range where we can track the relaxation rate (2~K~--~6~K). This means that the temperature dependence of the dc resistivity is governed by the relaxation rate, thus any other relevant parameters that affect the dc resistivity, like charge carrier density and effective mass, do not depend on temperature in this range. This indicates that for this material the heavy-fermion state is already fully established at a temperature as high as 6~K, and there are no changes of Fermi surface volume for lower temperatures.

We want to point out that in our studies we have the full frequency dependence available for numerous temperatures; thus we can unambiguously determine the relaxation rate from the Drude response and follow it with increasing temperature.\cite{Scheffler2005c} This is in contrast to the optical conductivity spectra obtained previously on heavy fermions, where the Drude roll-off could never be fully observed and furthermore only a few temperatures were studied, and it is also in contrast to previous microwave experiments, where a detailed temperature dependence for the relaxation rate was obtained, but relying on the assumption that the difference between dc and microwave conductivity can be described within the Drude response.
Furthermore, we directly determine both real and imaginary parts of the conductivity. We can follow the Drude relaxation up to a temperature of 6~K; for higher temperatures the relaxation rate is higher than our accessible frequency range. This is unfortunate because the behavior of the Drude relaxation around both the antiferromagnetic transition at 14~K and the crossover between heavy-fermion and uncorrelated state around 100~K is of interest.

\subsection{Frequency-dependent relaxation rate\label{SecFreqDepRelRate}}

\begin{figure}
\resizebox{1.00\columnwidth}{!}{%
  \includegraphics{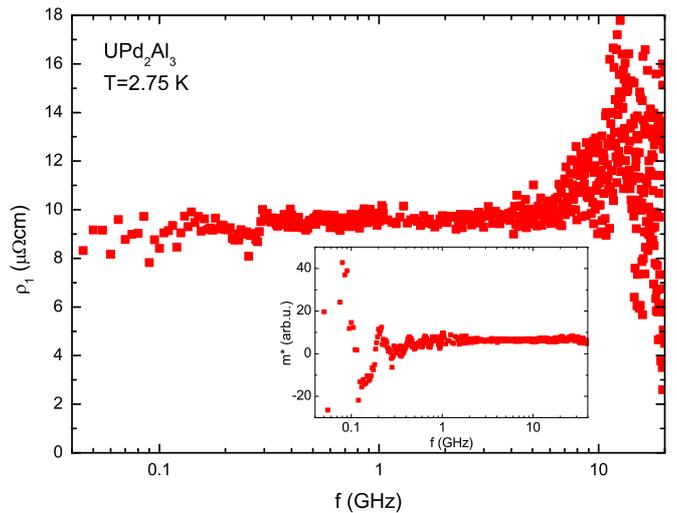}
}
%\vspace{5cm}       % Give the correct figure height in cm
\caption{(Color online) Real part of the complex resistivity (proportional to frequency-dependent relaxation rate) of UPd$_2$Al$_3$ at 2.75K. Inset: frequency-dependent effective mass (in arbitrary units).}
\label{FigExtendedDrude}       % Give a unique label
\end{figure}

For our compound, a quadratic temperature dependence of the dc resistivity is observed between $T_c$ and approximately 3.2~K, as shown in the inset of Fig.\ \ref{FigSamplesRdc} and known from literature.\cite{Huth1993,Huth1994b,Hiroi1997} This can be described as $\rho_{\rm dc}=\rho_0 + A T^2$, where $A =2.3\cdot 10^{-7}\ {\rm \Omega}$cm/K$^2$, similar to previous studies.\cite{Dalichaouch1992,Bakker1993} Such a quadratic temperature dependence is usually taken as an indicator for Fermi-liquid behavior, which should lead to a characteristic frequency dependence of the relaxation rate as well.\cite{Degiorgi1999}
But our microwave conductivity measurements can be described well by a simple Drude response, i.e. assuming a frequency-independent relaxation rate. Thus any possible frequency dependence of the relaxation rate has to be very small; in fact it is, as we will show, not resolvable with our present spectrometer. However, since our experiment for the first time at all allows studying the frequency dependence of the relaxation rate at microwave frequencies, we present our analysis in the following.

According to the extended Drude model as discussed in Section \ref{FreqDepRelRate}, one can use Eq.\ (\ref{EqFreqDepGamma}) to determine the frequency-dependent relaxation rate. To obtain absolute values of the relaxation rate, one has to include a prefactor (the unrenormalized plasma frequency), which is not known {\it a priori}, and furthermore is assumed to be temperature independent. To avoid these complications, in Fig.\ \ref{FigExtendedDrude} we plot the real part of the frequency-dependent resistivity $\rho(f)$, which is proportional to the frequency-dependent relaxation rate if those assumptions hold.
As seen in Fig.\ \ref{FigExtendedDrude}, the real part of the resistivity (and thus the relaxation rate) is constant for frequencies below 8~GHz, as expected from the perfect Drude behavior in the conductivity spectra. For higher frequencies, our data of $\rho_1(f)$ show a frequency dependence. However, we attribute this effect to the errors in our experiment, related to the achieved reproducibility of experimental conditions that we need for the calibration procedure: with increasing frequency, i.e. decreasing wavelength, it becomes harder to reproduce the damping and the phase shift introduced by the coaxial cable. Since we here discuss data at frequencies above 8~GHz, i.e. more than double the relaxation rate, the conductivity is mostly imaginary [$\sigma_2(f) > \sigma_1(f)$], and thus the phase of the measured reflection coefficient is much more relevant for the obtained values of $\rho_1$ than at lower frequencies, making the calibration particularly challenging.

Fermi-liquid theory predicts for the relaxation rate - and the real part of the resistivity - a quadratic dependence on both temperature and frequency, $\rho_1 (T,\omega) = \rho_0 + AT^2 + B\omega^2$. From the dc resistivity measurements, we know $\rho_0$ and $A$ for our sample, but since the prefactor $B$ of the frequency-dependent term (or equivalently the ratio $A/B$) depends on the particular Fermi surface \cite{Rosch2005,Rosch2006} and is not known for the case of UPd$_2$Al$_3$, we cannot calculate any Fermi-liquid prediction for our frequency-dependent data at this point. If instead we use the frequently stated generic number $(2 \pi)^2$ for $A/B$, \cite{Degiorgi1999} we can calculate a frequency-dependent resistivity, but the resulting frequency dependence is much smaller than our experimental resolution for the data in Fig.\ \ref{FigExtendedDrude}.
The very small contribution of a possible frequency-dependent relaxation rate also explains why we can describe our conductivity spectra with a simple Drude formula: at low temperatures, the dominant scattering contributions are connected to the residual resistivity (due to defects) and the $T^2$-term, both of which are independent of frequency. As a result, the total relaxation rate is also independent of frequency, leading to the simple Drude description.

We want to mention that also previous optical and microwave studies on heavy fermions tried to observe the Fermi-liquid contribution, but were not conclusive.\cite{Sulewski1988,Tran2002}
How could one observe the so-far elusive quadratic frequency dependence of relaxation rate and resistivity as predicted by Fermi liquid theory? In general, heavy-fermion materials are good candidates for such an experiment because the prefactors $A$ and $B$ scale quadratically with the effective mass, i.e. they should be strongly enhanced for heavy fermions compared to normal metals.\cite{Kadowaki1986} But for the present case of the microwave conductivity of UPd$_2$Al$_3$, this is still not sufficient, because the frequency-in\-de\-pen\-dent contributions $\rho_0 + AT^2$ are much stronger than the frequency-dependent one $B\omega^2$. One possible experimental approach would be to work at higher frequencies, but our experiment already reaches the present limits for broadband microwave spectroscopy at cryogenic temperatures. Furthermore, going to higher frequencies might not reveal a Fermi liquid response, because additional contributions to the frequency-dependent conductivity, which go beyond the Drude or Fermi-liquid response of the heavy conduction electrons, are know to occur at frequencies as low as 60 GHz.\cite{Dressel2002b,Dressel2002c} Another approach would be to reduce the frequency-independent contribution, by either reducing the temperature or the residual resistivity. Both are not possible for the present experiment, as we already employ thin film samples of the highest quality and we cannot reduce the temperature because UPd$_2$Al$_3$ becomes superconducting at 2.0~K. Thus, the most viable strategy to observe the frequency-squared Fermi-liquid prediction for the relaxation rate is to study materials with an even higher effective mass and to study metallic (non-superconducting) samples of very high quality at very low temperature.

Using the extended Drude analysis, we can also obtain the frequency dependence of the effective mass. Again, since we do not know the precise value of the unrenormalized plasma frequency, we only calculate relative values. The result is shown in the inset of Fig.\ \ref{FigExtendedDrude}. Within our experimental resolution, we cannot detect any frequency dependence of the effective mass. This is consistent with previous studies on UPd$_2$Al$_3$,\cite{Dressel2002b,Dressel2002c} where deviations from the enhanced, low-frequency mass were inferred only for frequencies above the spectral range addressed here.
The large data scattering in the effective mass at low frequencies is due to the fact that for these frequencies the imaginary part of the conductivity is basically zero, and therefore the obtained effective mass, following Eq.\ (\ref{EqFreqDepMstar}), is directly proportional to the error of the measurement and the calibration procedure.

As mentioned above, a quantitative extended Drude analysis requires knowledge of the {\it unrenormalized} plasma frequency $\omega_p$, which we unfortunately lack here. But we can use our microwave data to determine the {\it renormalized} plasma frequency $\omega_p^*$ at low temperatures.\cite{Degiorgi1997} 
Rewriting Eq.\ (\ref{EqFreqDepGamma}) for the renormalized quantities, we obtain $\Gamma^*(\omega) = \omega_p^{* 2} /(4\pi) \rho_1(\omega)$, which we can apply to our data of $\rho_1=9 \times 10^{-6}\ {\rm \Omega}$cm (see Fig.\ \ref{FigExtendedDrude}) and $\tau^*=1/\Gamma^* = 4.8 \times 10^{-11}$~s (from Drude fit). As expected, the resulting renormalized plasma frequency, $\omega_p^* /(2\pi c) = 840$~cm$^{-1}$ is considerably smaller than previous estimates from optical data where the actual Drude response could not be observed.\cite{Degiorgi1997}

\subsection{Sample dependence\label{SecSampleDep}}

The data and analysis presented so far was obtained on a single, high-quality sample (sample 1). Experiments on additional samples lead to consistent results, as discussed now for two of them. These samples are prepared in the same way as sample 1, except for the film thickness (samples 2 and 3 are 40~nm thick, whereas sample 1 is 150~nm thick). The quality of the sample clearly governs the low-temperature properties: 
Fig.\ \ref{FigSamplesRdc} shows the dc resistivity as a function of temperature for samples 1, 2, and 3. While the resistivity is roughly the same for high temperatures and the kink at the N\'eel temperature is observable for all samples, the residual resistivity and the transition to the superconducting state depend on sample quality. Sample 2, with a residual resistivity ratio of 4.4 becomes superconducting below 1.0~K whereas for sample 3, with even smaller resistivity ratio of 2.8, no superconductivity is expected even for lowest temperatures.

\begin{figure}
\resizebox{1.00\columnwidth}{!}{%
  \includegraphics{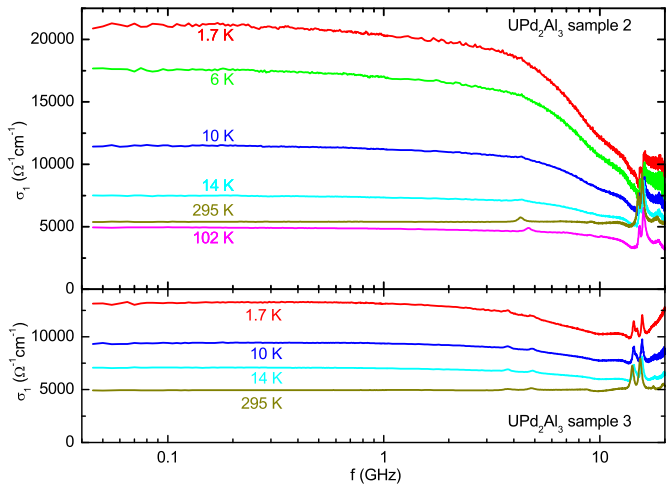}
}
%\vspace{5cm}       % Give the correct figure height in cm
\caption{(Color online) Real part of the frequency-dependent conductivity of UPd$_2$Al$_3$ samples 2 and 3 for a set of different temperatures.}
\label{FigUPA001And002Spectra}       % Give a unique label
\end{figure}

In Fig.\ \ref{FigUPA001And002Spectra}, we present conductivity spectra for samples 2 and 3 for a set of temperatures. For both samples, we observe a frequency-independent conductivity at high temperature, as expected and known from sample 1 (shown in Fig.\ \ref{FigSpectraEtc}). 
For low temperatures, we find a Drude roll-off in the real part of the conductivity also for samples 2 and 3, but at higher frequencies compared to sample 1 (Fig.\ \ref{FigSpectraEtc}; fit results for $\tau^*$ at 1.7~K are $1.4 \times 10^{-11}$~s for sample 2 and $9.5 \times 10^{-12}$~s for sample 3). In fact, for the lowest temperature of 1.7~K, we can only observe a reduction of $\sigma_1(f$=13~GHz) compared to $\sigma_0$ by 50\% for sample 2 and by 25\% for sample 3. This is consistent with the reduction of $\sigma_0$ of these samples compared to sample 1. 
Thus, the low-temperature microwave conductivity spectra of samples 2 and 3 also display the Drude response of the heavy fermions, but due to the enhanced residual resistivity compared to sample 1, the relaxation rate is higher and the Drude roll-off shifted to higher frequencies.

From the experimental point of view, one particular problem of the measurements of samples 2 and 3 are the pronounced resonances in the conductivity spectra for frequencies above 13~GHz. These resonances stem from standing waves in the dielectric substrate and the sample holder, and they become more pronounced for more resistive samples. For the thicker sample 1, these resonances are much less pronounced. On the other hand, the experimental advantage of the higher resistivity of samples 2 and 3, compared to sample 1, is the enhanced sensitivity of the spectrometer,\cite{Scheffler2005a} as evident from the reduced \lq noise\rq{} in the conductivity spectra.

\section{Conclusions and outlook}

Applying a broadband microwave spectrometer to thin film samples of UPd$_2$Al$_3$, we are able to measure the microwave conductivity of this heavy-fermion material. While we observe conventional metallic behavior at high temperature, the low-temperature data reveal the full Drude response of the heavy charge carriers. These experiments clearly establish that the heavy fermions in UPd$_2$Al$_3$ follow an extremely slow Drude relaxation, at frequencies that are much lower than for normal metals. From the spectra we can directly determine the relaxation rate and find that, as a function of temperature, it scales with the dc resistivity. Further evidence for this relation between Drude relaxation rate and dc resistivity is the effect of residual resistivity, as studied via sample dependence.

Our studies on UPd$_2$Al$_3$ show that the heavy-fermion relaxation rate in this material is considerably lower than assumed previously.\cite{Dressel2002b,Dressel2002c} An open question now is wheth\-er this holds for heavy fermions in general or is a unique feature of UPd$_2$Al$_3$.\cite{Scheffler2009} 
First experiments on thin films of UNi$_2$Al$_3$ have revealed similar Drude behavior in the GHz range,\cite{Scheffler2006} but more detailed investigations on this compound are necessary before final conclusions can be drawn. An aspect of particular interest here is the anisotropy of the microwave conductivity: a modification of the present spectrometer allows for the study of anisotropic materials if they are available as thin films with the anisotropy in the plane.\cite{Scheffler2007} Such studies of anisotropic optical properties are under way for UNi$_2$Al$_3$,\cite{Ostertag2010} and they could also be performed on $a^*$-oriented thin films of UPd$_2$Al$_3$.\cite{Foerster2007}
Additional studies of the microwave conductivity on other heavy-fermion materials, in particular based on Ce or Yb instead of U, would be desired to demonstrate whether the extremely low relaxation is generic for heavy-fermion compounds. If materials with even higher effective masses could be studied with microwaves, a shift of the Drude response to yet lower frequencies might be observed.
Higher effective masses might also enable the still missing observation of clear Fermi-liquid behavior in the optical conductivity of a metal.

\thanks{We thank Dimitri Basov, Achim Rosch, and Andrew Schofield for helpful discussions. This work was supported by the DFG.}%

%
% BibTeX users please use
% \bibliographystyle{}
% \bibliography{}

\begin{thebibliography}{}
\bibitem{Grewe1991}N. Grewe and F. Steglich, in {\it Handbook on the Physics and Chemistry of Rare Earths}, 
edited by K. A. Gschneidner and L. Eyring (Elsevier, Amsterdam, 1991), Vol. 14, p. 343.
%
\bibitem{Dressel2002a}M. Dressel and G. Gr\"{u}ner,
{\it Electrodynamics of Solids} (Cambridge University Press, Cambridge, 2002).
%
\bibitem{Degiorgi1999}L. Degiorgi,
Rev. Mod. Phys. \textbf{71}, 687 (1999).
%The electrodynamic response of heavy-electron compounds
%
\bibitem{Huth1993}M. Huth, A. Kaldowski, J. Hessert, Th. Steinborn, and H. Adrian,
Solid State Commun. \textbf{87}, 1133 (1993).
%Preparation and Characterization of Thin Films of the Heavy Fermion Superconductor UPd$_2$Al$_3$
%
\bibitem{Huth1994}M. Huth, A. Kaldowski, J. Hessert, C. Heske, and H. Adrian,
Physica B \textbf{199\&200}, 116 (1994).
%UPd$_2$Al$_3$ heavy fermion superconducting films

\bibitem{Scheffler2005a}M. Scheffler and M. Dressel,
Rev. Sci. Instrum. \textbf{76}, 074702 (2005).
%Broadband microwave spectroscopy in Corbino geometry for temperatures down to 1.7 K

\bibitem{Scheffler2005b}M. Scheffler, M. Dressel, M. Jourdan, and H. Adrian,
Physica B \textbf{359-361}, 1150 (2005).
%Direct observation of Drude behavior in the heavy-fermion UPd2Al3 by broadband microwave spectroscopy

\bibitem{Scheffler2005c}M. Scheffler, M. Dressel, M. Jourdan, and H. Adrian,
Nature \textbf{438}, 1135 (2005).
%Extremely slow Drude relaxation of correlated electrons

\bibitem{Drude1900}P. Drude, {Phys. Z.} \textbf{1}, 161 (1900).

\bibitem{Dressel2006}M. Dressel and M. Scheffler,
Ann. Phys. (Leipzig) \textbf{15}, 535 (2006).
%Verifying the Drude response

\bibitem{Varma1985}C.M. Varma,
Phys. Rev. Lett {\bf 55}, 2723 (1985).
%Phenomenologial Aspects of Heavy Fermions

\bibitem{Millis1987}A.J. Millis and P.A. Lee,
Phys. Rev. B \textbf{35}, 3394 (1987).
%Large-orbital-degeneracy expansion for the lattice Anderson model

\bibitem{Sulewski1988}P.E. Sulewski, A.J. Sievers, M.B. Maple, M.S. Torikachvili, J.L. Smith, and Z. Fisk,
Phys. Rev. B \textbf{38}, 5338 (1988).
%Far-infrared absorptivity of UPt$_3$

\bibitem{Beyermann1988a}W.P. Beyermann, G. Gr\"uner, Y. Dalichaouch, and M.B. Maple,
Phys. Rev. Lett \textbf{60}, 216 (1988).
%Relaxation-Time Enhancement in the Heavy-Fermion System CePd$_3$

\bibitem{Awasthi1993}A.M. Awasthi, L. Degiorgi, G. Gr\"uner, Y. Dalichaouch, and M.B. Maple,
Phys. Rev. B \textbf{48}, 10692 (1993).
%Complete optical spectrum of CePd$_3$

\bibitem{Degiorgi1997}L. Degiorgi, St. Thieme, H.R. Ott, M. Dressel, G. Gr\"uner, Y. Dalichaouch, M.B. Maple, Z. Fisk, C. Geibel, and F. Steglich,
Z. Phys. B \textbf{102}, 367 (1997).
%The electrodynamic response of heavy-electron materials with magnetic phase transitions

\bibitem{Tran2002}P. Tran, S. Donovan, and G. Gr\"uner,
Phys. Rev. B \textbf{65}, 205102 (2002).
%Charge excitation spectrum in UPt$_3$

\bibitem{Scheffler2009}M. Scheffler, M. Dressel, and M. Jourdan,
J. Phys. Conf. Ser. \textbf{150}, 042174 (2009).
%Low-temperature microwave response of heavy-fermion compounds

\bibitem{Dordevic2001}S.V. Dordevic, D.N. Basov, N.R. Dilley, E.D. Bauer, and M.B. Maple,
Phys. Rev. Lett. \textbf{86}, 684 (2001).
%Hybridization Gap in Heavy-Fermion Compounds

\bibitem{Singley2002}E.J. Singley, D.N. Basov, E.D. Bauer, and M.B. Maple,
Phys. Rev. B \textbf{65}, 161101(R) (2002).
%Optical conductivity of the heavy fermion superconductor CeCoIn$_5$

\bibitem{Mena2005}F.P. Mena, D. van der Marel, and J.L. Sarrao,
Phys. Rev. B \textbf{72}, 045119 (2005).
%Optical conductivity of the CeMIn$_5 (M=Co,Rh,Ir)$

\bibitem{Okamura2007}H. Okamura, T. Watanabe, M. Matsunami, T. Nishihara, N. Tsujii, T. Ebihara, H. Sugawara, H. Sato, Y. \=Onuki, Y. Isikawa, T. Takabatake, and T. Nanba,
J. Phys. Soc. Jpn. \textbf{76}, 023703 (2007).
%Universal Scalind in the Dynamical Conductivity of Heavy Fermion Ce and Yb Compounds

\bibitem{Rosch2005}A. Rosch and P.C. Howell,
Phys. Rev. B \textbf{72}, 104510 (2005).
%Zero-temperature optical conductivity of ultraclean Fermi liquids and superconductors

\bibitem{Rosch2006}A. Rosch,
Ann. Phys. (Leipzig) \textbf{15}, 526 (2006).
%Optical conductivity of clean metals

\bibitem{Geibel1991}C. Geibel, C. Schank, S. Thies, H. Kitazawa, C.D. Bredl, A. B\"ohm, M. Rau, A. Grauel, R. Caspary, R. Helfrich, U. Ahlheim, G. Weber, and F. Steglich,
Z. Phys. B \textbf{84}, 1 (1991).
%Heavy-fermion superconductivity at T$\rm _c$=2K in the antiferromagnet UPd$_2$Al$_3

\bibitem{Jourdan1999}M. Jourdan, M. Huth, and H. Adrian,
Nature \textbf{398}, 47 (1999).
%Superconductivity mediated by spin flucturations in the heavy-fermion compound UPd$_2$Al$_3$

\bibitem{Sato2001}N.K. Sato, N. Aso, K. Miyake, R. Shiina, P. Thalmeier, G. Varelogiannis, C. Geibel, F. Steglich, P. Fulde, and T. Komatsubara,
Nature \textbf{410}, 340 (2001).
%Strong coupling between local moments and superconducting \lq heavy\rq electrons in UPd$_2$Al$_3$

\bibitem{Dressel2002b}M. Dressel, N. Kasper, K. Petukhov, B. Gorshunov, G. Gr\"uner, M. Huth, and H. Adrian,
Phys. Rev. Lett. \textbf{88}, 186404 (2002).
%Nature of Heavy Quasiparticles in Magnetically Ordered Heavy Fermions UPd$_2$Al$_3$ and UPt$_3$

\bibitem{Dressel2002c}M. Dressel, N. Kasper, K. Petukhov, D.N. Peligrad, B. Gorshunov, M. Jourdan, M. Huth, and H. Adrian,
Phys. Rev. B \textbf{66}, 035110 (2002).
%Correlation gap in the heavy-fermion antiferromagnet UPd$_2$Al$_3$


\bibitem{RemarkSensitivityBulk}Estimating from the dc conductivity, one obtains a difference in reflection coefficient of 0.0004 between a bulk UPd$_2$Al$_3$ sample and the reference of a perfect short. This difference would by far not be resolvable.

\bibitem{Steinberg2008}K. Steinberg, M. Scheffler, and M. Dressel,
Phys. Rev. B \textbf{77}, 214517 (2008).
%Quasiparticle response of superconducting aluminum to electromagnetic radiation


\bibitem{RemarkAnomalousSkinEffect}Since we operate in the thin-film limit, we do not have to consider the skin effect at all; in particular it is not relevant whether our sample material (in bulk) exhibits the normal or anomalous skin effect, a distinction one has to make when studying high-quality bulk samples of metals at cryogenic temperatures. But we point out that although we study high-quality samples with extremely low relaxation rates, the bulk material will not enter the anomalous skin effect regime: the mean free path of approximately 60~nm at low temperatures is not extremely long, and thus the electrodynamic response of the electrons is local.


\bibitem{Huth1994b}M. Huth, J. Hessert, M. Jourdan, A. Kaldowski, and H. Adrian,
Phys. Rev. B \textbf{50}, 1309 (1994).
%Coherence effects in the low-temperature Hall coefficient of the heavy-fermion system UPd$_2$Al$_3$

\bibitem{Hiroi1997}M. Hiroi, M. Sera, N. Kobayashi, Y. Haga, E. Yamamoto, and Y. \=Onuki,
J. Phys. Soc. Jpn. \textbf{66}, 1595 (1997).
%Thermal Conductivity of a Heavy Fermion Superconductor UPd$_2$Al$_3$ Single Crystal

\bibitem{Dalichaouch1992}Y. Dalichaouch, M.C. de Andrade, and M.B. Maple,
Phys. Rev. B \textbf{46}, 8671 (1992).
%Superconducting and magnetic properties of the heavy-fermion compounds U$T_2$Al$_3$ ($T$=Ni,Pd)

\bibitem{Bakker1993}K. Bakker, A. de Visser, L.T. Tai, A.A. Menovsky, and J.J.M. Franse,
Solid State Commun. \textbf{86}, 497 (1993).
%The Effect of Pressure on the Antiferromagnetic and Superconducting States of the Heavy-Fermion UPd$_2$Al$_3$


\bibitem{Shoenberg1984}D. Shoenberg, 
{\it Magnetic oscillations in metals} (Cambridge University Press, Cambridge, 1984).

\bibitem{Inada1999}Y. Inada, H. Yamagami, Y. Haga, K. Sakurai, Y. Tokiwa, T. Honma, E. Yamamoto, Y. \=Onuki, and T. Yanagisawa,
J. Phys. Soc. Jpn. \textbf{68}, 3643 (1999).
%Fermi Surface and de Haas-van Alphen Osciallation in both the Normal and Superconducting Mixed States of UPd$_2$Al$_3$


\bibitem{Kadowaki1986}K. Kadowaki and S. B. Woods,
Solid State Commun. \textbf{58}, 507 (1986).
%Universal relationship of the resistivity and specific heat in heavy-fermion compounds


\bibitem{Scheffler2006}M. Scheffler, M. Dressel, M. Jourdan,  and H. Adrian,
Physica B \textbf{378-380}, 993 (2006).
%Dynamics of heavy fermions: Drude response in UPd$_2$Al$_3$ and UNi$_2$Al$_3$

\bibitem{Scheffler2007}M. Scheffler, S. Kilic, and M. Dressel,
Rev. Sci. Instrum. \textbf{78}, 086106 (2007).
%Strip-shaped samples in a microwave Corbino spectrometer

\bibitem{Ostertag2010}J.P. Ostertag, M. Scheffler, M. Dressel, and M. Jourdan,
Phys. Status Solidi B (in press), doi:10.1002/pssb.200983030.
%Anisotropic optics on heavy fermions

\bibitem{Foerster2007}M. Foerster, A. Zakharov, C. Herbort, and M. Jourdan,
J. Magn. Magn. Mater. \textbf{310}, 346 (2007).
%Electronic properties of a*-oriented UPd$_2$Al$_3$ thin films




\end{thebibliography}

\end{document}